\documentclass[11pt]{article}
\usepackage{graphicx} \textheight 22.5cm\topmargin
-0.2 in \textwidth 15.5cm\oddsidemargin 0cm\evensidemargin 0cm
\long\def\symbolfootnote[#1]#2{\begingroup
\def\thefootnote{\fnsymbol{footnote}}\footnote[#1]{#2}\endgroup}
\usepackage{indentfirst}
\usepackage{amsmath}
\usepackage{stmaryrd}
\usepackage{bm}
\usepackage{slashbox}

\begin{document}
\title{Possible S-wave Bound-States of Two Pseudoscalar Mesons }
\author{Yin-Jie Zhang$^1$, Huan-Ching Chiang$^{2,1}$, Peng-Nian Shen$^{3,1}$
and Bing-Song Zou$^{1,4}$\footnote{zoubs@mail.ihep.ac.cn}\\
$^1$ Institute of High Energy Physics, P.O.Box 918(4), Beijing 100049, China\\
$^2$ South-West University, Chongqing 400715, China\\
$^3$ CCAST (World Lab.), P.O.Box 8730, Beijing 100080, China\\
$^4$ Center of Theoretical Nuclear Physics, National Laboratory of
Heavy Ion Accelerator,\\ Lanzhou 730000, China}

\date{April 28, 2006}
\maketitle
\begin{abstract}
Using the potentials derived from vector-meson-exchanges,  the
$K\bar K$, $DK$, $B\bar K$, $D\bar D$, $B\bar B$, $BD$ , $\bar D
K$, $BK$ and $B\bar D$ systems are studied.  Possible S-wave
bound-states of two pseudoscalar mesons are discussed. We find
that systems of $DB$, $D\bar D$, $B\bar B$ and $DK$ with isospin 0
are likely to form S-wave bound states via strong interactions
while the others are unlikely to be bound by the strong
interaction sector alone. With the Coulomb potential, the $K^+K^-$
system can form an atomic bound state - the kaonium. The influence
of the one-meson-exchange potential on the ground state energy of
the kaonium and its decay widths to $\pi\pi$ and $\pi\eta$ are
evaluated.
\end{abstract}

\section{Introduction}

Since the 1970's, the possible existence of the nuclear-like bound
states of mesons has been one of the interesting subjects in
hadron physics. Some hardrons such as the $\psi(4040)$
\cite{psi4040}, $\eta(1440)$ \cite{tornqvist1994}, $f_1(1420)$
\cite{f1420}, $f_J(1720)$ \cite{f1720}, and the $f_0(980)$ and
$a_0(980)$
\cite{weinstein1990,weinstein1982,close1993,ahcasov1997} were
explained as the nuclear-like bound states of mesons. T$\ddot{\rm
o}$rnqvist \cite{Tornqvist1991} argued that the one-pion-exchange
potential is likely to form a few states composed of two
ground-state mesons. In 2003 the Belle Collaboration reported a
new narrow charmonium state at $3872\pm0.6(stat)\pm0.5(syst)$ MeV
and with a width $<2.3$MeV(95\% C.L.), which has been confirmed by
several experiments \cite{choi,acosta,abazov,aubert}. Almost
immediately, T$\ddot{\rm o}$rnqvist \cite{tornqvist03} claimed
that this state is the one he predicted long ago. Indeed the
proximity of the X to $D^0\bar D^{0*}$ threshold led to a
speculation that the X is a $D\bar D^*$ resonance
\cite{close04,wong05,Swanson04}.

Recently, beginning  with the discovery of $D_s(2317)$, more than
10 heavy mesons are reported, such as the $B_c$, $h_c$, $\eta_c'$,
$D_s(2460)$, $X(3872)$, $X(3940)$, $Y(3940)$, $Z(3930)$ and
$Y(4260)$. The new findings have generated much enthusiasm for
understanding the nature of the new mesons.  A survey of the
experimental, phenomenological and theoretical status of the new
heavy mesons can be found in a recent review article by Swanson
\cite{swanson0601110}. Different models are used to study the
spectroscopy. For example, Liu, Zeng and Li \cite{liuzeng}
suggested that the Y(4260) is a $\chi_{c1}\rho$ molecule state
bound by the $\sigma$ exchange. Barnes, Close and Lipkin
\cite{barnesclose} have suggested that the $D_{s0}$ and $D_{s1}$
may be $DK^{(*)}$ molecules. However, until now, none of these
states is well-established as a molecular state.

The existence of so many controversial states strongly suggest
possible two-meson structure. However, the dynamics of bound
states of two mesons is not well-understood. It is worthy to study
systems of two mesons. For example, the understanding for the
structure of the $f_0(980)$ has been controversial for many years.
It might be a $q\bar q$ state, a $q^2 \bar{q}^2$ state, or a
$K\bar K$ molecule state. Krewald et al. \cite{1} studied the
kaon-antikaon system by using the strong interactions  generated
from vector-meson-exchange in the frame work of the SU(3)
invariant effective Lagrangian. They have shown that one-meson
exchange potentials derived from this Lagrangian in the
non-relativistic limit are sufficient to bind $K\bar K$ into a
kaonic molecule with a mass and decay width that closely match the
experimental values of the $f_0(980)$ meson. However, in their
study the momentum dependent terms of potentials  are neglected
and a quite large cut-off parameter of the form factor is used.

In present paper, we extend the study to $K\bar K$, $DK$, $B\bar
K$, $D\bar D$, $B\bar B$, $BD$ , $\bar DK$, $BK$, $B\bar D$
systems by including the momentum dependent terms of
one-vector-meson exchange interactions. The momentum dependent
terms result non-local potentials of two mesons in coordinate
space.  We use these nonlocal potentials to search for possible
bound states of two-meson systems.

The model is briefly described in a generalized form in section 2.
In section 3 we discuss possible bound states of the two-meson
systems.  The $K\bar K$ system is discussed in detail. Both strong
and Coulomb potential are included in the calculation. The strong
interaction and its influence on the ground state energy and decay
widths of the kaonium are discussed. We also extend our
calculations to some other systems including heavy B and D mesons,
where some possible S-wave bound states are found. Finally we give
our summary and discussion in section 4.

\section{The Model}

In order to investigate the possible bound states of the $K\bar
K$, $DK$, $B\bar K$, $D\bar D$, $B\bar B$, $BD$, $KK$, $\bar DK$,
$BK$, $DD$, $BB$ and $B\bar D$ systems in the framework of
non-relativistic Schr\"odinger equation, we first derive the
interaction potentials between two pseudoscalar mesons from
vector-meson exchange diagrams which are found to be the dominant
t-channel exchange interactions \cite{1,Lohse1990,wufq}. The
relevant interaction Lagrangian of the
pseudoscalar-pseudoscalar-vector coupling can be written in the
following form,
\begin{equation}
\mathcal L_{PPV}=-\frac12iG_VTr([P,\partial_\mu P]V^\mu).
\end{equation}
Usually P and V stand for the fields of pseudoscalar and vector
octets, respectively. $P$ and $V$ are $3\times 3$ matrices,
\begin{equation}
P=\sqrt2 \left(
\begin{array}{ccc}
\frac{1}{\sqrt{2}}\pi^0 + \frac{1}{\sqrt{6}}\eta_8 & \pi^+ & K^+\\
\pi^- & -\frac{1}{\sqrt{2}}\pi^0 + \frac{1}{\sqrt{6}}\eta_8 & K^0\\
K^- & \bar{K}^0 & - \frac{2}{\sqrt{6}}\eta_8
\end{array}
\right),
\end{equation}
and
\begin{equation}
V = \sqrt2\left(
\begin{array}{ccc}
\frac{1}{\sqrt{2}}\rho^0 + \frac{1}{\sqrt{2}}\omega & \rho^+ & K^{*+}\\
\rho^- & -\frac{1}{\sqrt{2}}\rho^0 + \frac{1}{\sqrt{2}}\omega & K^{*0}\\
K^{*-} & \bar{K}^{*0} & \phi
\end{array}
\right).
\end{equation}

For a special case of the $K\bar K$ system the relevant
Lagrangians are
$$\mathcal L_{KK\rho}=iG_V\{[\bar K\vec \tau(\partial_\mu
K)-(\partial_\mu \bar K)\vec\tau K]\cdot \vec\rho^\mu\}$$
$$\mathcal L_{KK\omega}=iG_V [\bar
K(\partial_\mu K)-(\partial_\mu \bar K)K]\omega^\mu$$ $$\mathcal
L_{KK\phi}=-\sqrt2iG_V[\bar K(\partial_\mu K)-(\partial_\mu \bar
K)K]\phi^\mu$$ The $K$ and $\bar K$ is
$$K=\left(\begin{array}{c}K^+\\ K^0\end{array}\right)~~~~,~~~~\bar
K=(K^-~,~\bar K^0)$$

Here $G_V$ is the coupling constant,which can be fixed in terms of
the experimental value of $\rho \pi \pi $ coupling constant
$g_{\pi \pi\rho}=2G_V$.

In present paper we  deal with systems of two pseudoscalar mesons
including heavy pseudoscalar mesons $D$ and $B$. We will
generalize Eq.(1) to include the $D$ and $B$ pseudoscalar  mesons.
In this case, the strengths of their couplings to $\rho$, $\omega$
and $\phi$ mesons can be determined from the quark model.

By evaluating the meson-meson scattering amplitude in the Born
approximation from the  t-channel vector meson exchange the
interaction potentials of two mesons in momentum space can be
obtained. In the case of the $K\bar K$ system, one finds
$$V(M_V,\vec
p,\vec{q})=-\frac{g_{KKV}^2}{4M_K^2}C_I\frac{1}
{\vec{q}^2+M_V^2}(4M_K^2+8\vec{p}^2+8\vec{q}
\cdot\vec{p}+3\vec{q}^2).$$
Here subscript $V$ stands for the
vector meson $\rho$, $\omega$ or $\phi$. $M_V$ is the mass of the
V meson. $M_K$ is the mass of the kaon, $\vec p$ is the momentum
of the kaon in the center-of-mass frame. $\vec q$ is the
3-momentum transfer between $K$ and $\bar K$ in the t-channel. The
relevant coupling constants $g_{KKV}$ are related by SU(3)
symmetry relations,

\begin{equation}
g_{KK\rho}=G_V~,~g_{KK\omega}=G_V ~,~g_{KK\phi}=-\sqrt2G_V .
\end{equation}
$C_I$ is the isospin factor with I=0, 1. For the exchange of
$\rho$, $\omega$, $\phi$, it is respectively
$$C_0=\left\{\begin{array}{rl}3&~~{\rm for}~\rho\\1&~~{\rm for}~\omega
\\1&~~{\rm for}~\phi\end{array}\right.~~;~~C_1=\left\{\begin{array}{rl}-1&~~{\rm for}~\rho\\1&~~{\rm for}~\omega
\\1&~~{\rm for}~\phi\end{array}\right.$$

For each vector-meson exchange, after performing a Fourier
transformation of the scattering amplitude the interaction
potential in coordinate space can be obtained as
\begin{equation}
V(M_V,\vec{r}) =-\frac{g_{KKV}^2}{4\pi}C_I\left[U(M_V,r)
\frac{2\vec{p}^2}{M_K^2}-\displaystyle\frac{2i}{M_K^2}\nabla
U(M_V,r)
\cdot\vec{p}-\frac{3}{4M_K^2}\nabla^2U(M_V,r)+U(M_V,r)\right]
\end{equation}
with
$$\begin{array}{rcl}
 U(M,r)&=&4\pi\displaystyle\int \frac{d^3\vec{k}}{(2\pi)^3}\frac{[F^t(\vec{k})]^2}
 {M^2+\vec{k}^2}e^{i\vec{k}\cdot\vec{r}}\cr\noalign{\vskip2truemm}
 &=&\displaystyle\frac{e^{-Mr}}{r}-\frac{e^{-\Lambda
 r}}{r}\left[1+\frac{1}{16}\left(11-\frac{4M^2}{\Lambda^2}+\frac{M^4}{\Lambda^4}
 \right)\left(1-\frac{M^2}{\Lambda^2}\right)(\Lambda r)\right.\cr\noalign{\vskip2truemm}
 &&+\displaystyle\left.\frac{1}{16}\left(3-\frac{M^2}{\Lambda^2}\right)\left(1-\frac{M^2}{\Lambda^2}\right)^2
 (\Lambda r)^2+\frac{1}{48}\left(1-\frac{M^2}{\Lambda^2}\right)^3(\Lambda
 r)^3\right] .
 \end{array}
$$
Here, a form factor with a cutoff parameter $\Lambda$ for each
interaction vertex is added.
\begin{equation}
F^t(\vec{q})=\left(\frac{\Lambda^2-M^2}{\Lambda^2+\vec{q}^2}\right)^2
.
\end{equation}

The total potential of the $K\bar K$ system is a sum of
contributions from $\rho$, $\omega$ and $\phi$ exchanges.
\begin{equation}\label{potkk}
V({\vec r})=V(M_\rho,{\vec r})+V(M_\omega,{\vec r})+V(M_\phi,{\vec
r}).
\end{equation}

We assume that the coupling of the vector meson with two
pseudoscalar mesons originates from the coupling of the vector
meson with quarks inside the pseudoscalar mesons.  The $\rho$ and
$\omega$ mesons may couple to the  the $u$ and $d$ quarks, while
the $\phi$ meson only couples to the $s$ quark inside the
pseudoscalar meson. From this consideration and by using the quark
structure of the heavy mesons, we can generalize the derivation of
the vector-meson-exchange potential between $K\bar K$ to other
two-meson systems including $ DK$, $B\bar K$, $D\bar D$, $B\bar
B$, $BD$ , $\bar D K$, $BK$ and $B\bar D$. Because the $D$ and $B$
mesons do not contain the $s$ quark, the $\phi$ meson coupling
does not present. For $KK$ and $K\bar K$ systems $\rho$,$\omega$
and $\phi$  exchanges all contribute to the interaction. For
others systems containing the $D$ or $B$ meson there are only
contributions from $\rho$ and $\omega$ exchanges. For these
systems, the relevant interaction Lagrangians are written as

$$\mathcal L_{BB\rho}=iG_V\{[\bar B\vec\tau(\partial_\mu B)-(\partial_\mu\bar B)\vec\tau B]\cdot\vec\rho^\mu\}$$
$$\mathcal L_{BB\omega}=iG_V[\bar B(\partial_\mu B)-(\partial_\mu\bar B)B]\omega^\mu$$
$$\mathcal L_{DD\rho}=iG_V\{[D\vec\tau(\partial_\mu \bar D)-(\partial_\mu D)\vec\tau\bar D]\cdot\vec\rho^\mu\}$$
$$\mathcal L_{DD\omega}=iG_V[D(\partial_\mu\bar D)-(\partial_\mu
D)\bar D]\omega^\mu.$$ We assume that the coupling constant $G_V$ is
the same as the usual PPV coupling constant.
$$
\bar D=\left(\begin{array}{c}\bar{D^0}\\ D^-\end{array}\right)~;~B=\left(\begin{array}{c}B^+\\
B^0\end{array}\right)$$
$$D=
(D^0,D^+)~;~\bar B=(B^-,\bar B^0)$$
One can get the relevant
coupling constants from the Lagrangians
\begin{equation}
g_{DD\rho}=g_{BB\rho}=g_{KK\rho}=G_V~,~g_{DD\omega}=g_{BB\omega}=g_{KK\omega}=G_V
\end{equation}

From these Lagrangians  a non-relativistic Hamiltonian for the
two-pseudoscalar-meson system can be obtained as
\begin{equation}
H=[-\frac1{2\mu}+a(r)]\nabla^2+b(r)\frac\partial{\partial r}+c(r)
.
\end{equation}
Here $\mu$ stands for the reduced mass of the two mesons, and
\begin{equation}
a(r)=\sum\limits_V\frac{g_{PPV}^2}{4\pi}C_I\frac1{4m_1m_2}\left[2\frac{m_1^2+m_2^2}{m_1m_2}
+4\right]U(M_V,r),
\end{equation}
\begin{equation}
b(r)=\frac\partial{\partial r}a(r),
\end{equation}
\begin{equation}
c(r)=\sum\limits_V\frac{g_{PPV}^2}{4\pi}C_I\left[\frac1{4m_1m_2}\left(\frac{m_1^2+m_2^2}{m_1m_2}
+1\right)\left(\frac{\partial^2}{\partial r^2}+\frac
2r\frac{\partial }{\partial r}\right)U(M_V,r)-U(M_V,r)\right],
\end{equation}
with $m_1~,~m_2$  the masses of the pseudoscalar mesons.

$g_{PPV}$ is the pseudoscalar-pseudoscalar-vector coupling
constants, $C_I$ is an isospin factor. By using the quark
structure of mesons and comparing with the isospin factors for the
$K\bar K$ system we can get the isospin factors.

For $DK$,$B\bar K$, $D\bar D$, $ B\bar B$ and $ BD$ systems,
\begin{equation}
    C_0=\left\{\begin{array}{rl}3&~{\rm for}~\rho\\ 1&~{\rm
    for}~\omega\end{array}\right.~;\quad
    C_1=\left\{\begin{array}{rl}1&~{\rm for}~\rho\\ 1&~{\rm
    for}~\omega\end{array}\right. ~.
\end{equation}
For $\bar DK$, $BK$, $\bar D\bar D$, $ BB$ and $ B\bar
D$ systems,
\begin{equation}
     C_0=\left\{\begin{array}{rl}3&~{\rm for}~\rho\\ -1&~{\rm
     for}~\omega\end{array}\right.~;\quad
     C_1=\left\{\begin{array}{rl}-1&~{\rm for}~\rho\\ -1&~{\rm
     for}~\omega\end{array}\right.~.
\end{equation}

By solving the  Schr\"odinger equation with above Hamiltonian one
can search for  possible bound states of the two-meson systems.

There are two parameters in the model, the coupling constant
$g_{\rho\pi\pi}$ and the cutoff parameter $\Lambda$. The coupling
constant $g_{\rho\pi\pi}$ is determined by the decay width
$\Gamma(\rho\to\pi\pi)$ as $g_{\pi\pi\rho}^2/4\pi=2.8$. From
$G_V=g_{\pi\pi\rho}/2$ we get $G_V=3.0$. So the only free
parameter is the cutoff parameter $\Lambda$ for each system. In
general, the cutoff parameters can be different for different
pseudoscalar-meson systems. In the analysis of the $\pi\pi$,\ and
$\pi K$ phase shifts, the $\Lambda$ changes ranging from 1.5 GeV
to 2.0 GeV \cite{wufq}. We use the same values for all systems in
our study.

\section{Possible bound state}

We first discuss the $K\bar K$ system. The strong interaction
potential of this system is in the form of Eq.(\ref{potkk}). The
corresponding Schr\"odinger equation has a singular point which
depends on the reduced mass of two mesons to cause
$[-\frac1{2\mu}+a(r)]=0$ in Eq.(9). Because of the small mass of
the $K\bar K$ system and the existence of the singular point, the
one meson ($\rho,\omega,\phi$) exchange potential alone fails to
produce any $K\bar K$ bound state. By solving the Schr\"odinger
equation with Coulomb potential, one finds an atomic bound state
of the $K^+K^-$ system. The binding energy and the root mean
square radius of the atomic bound state are $\epsilon=6.58$ keV
and $<r^2>^{\frac12}=190.5\rm{fm}$, respectively. Combining the
Coulomb potential with the strong interaction potential derived
from vector-meson-exchanges, we find that the binding energy and
root mean square radius of the $K^+K^-$ atomic bound state are
changed to be $\epsilon=7.05\rm{keV}$ and
$<r^2>^{\frac12}=175.4\rm{fm}$, respectively. In the calculation
the cutoff parameter $\Lambda$ is taken to be a value of 2.0 GeV
which is in consistent with that used in Ref.\cite{wufq} and is
much smaller than that used in Ref. \cite{1} with
$\Lambda=4\sqrt{2}$ GeV. Even if we take $\Lambda=4\sqrt{2}$ GeV
or drop momentum-dependent terms but with $\Lambda=2$ GeV, we
cannot get $K\bar K$ bound state without Coulomb potential. Only
when we take $\Lambda=4\sqrt{2}$ GeV and drop the
momentum-dependent terms to be exactly the same as in Ref.
\cite{1}, we can get the strongly bound $K\bar K$ bound state as
in Ref. \cite{1}.

\begin{figure}\begin{center}
\setlength{\unitlength}{0.4cm}
\begin{picture}(11,4)
\put(0.5,0.5){\line(0,1){3}} \put(0.5,2){\line(1,0){4}}
\put(4.5,0.5){\line(0,1){3}}
\put(0.25,0){\makebox(0.5,0.3){$\bar{K}$}}
\put(0.25,3.5){\makebox(0.5,0.8){$\pi$}}\put(4.25,0){\makebox(0.5,0.3){$K$}}
\put(4.25,3.5){\makebox(0.5,0.8){$\pi$}}\put(1.8,2.25){\makebox(1.5,0.6){$K^*$}}
\put(8,0.5){\line(0,1){3}} \put(8,2){\line(1,0){4}}
\put(12,0.5){\line(0,1){3}}
\put(7.75,0){\makebox(0.5,0.3){$\bar{K}$}}
\put(7.75,3.5){\makebox(0.5,0.8){$\pi$}}\put(11.75,0){\makebox(0.5,0.3){$K$}}
\put(11.75,3.5){\makebox(0.5,0.8){$\eta$}}\put(9.3,2.25){\makebox(1.5,0.6){$K^*$}}
\end{picture}
\end{center}\caption{Decay diagrams of the $K\bar
K$ bound state to $\pi\pi$ and $\pi\eta$ via a $K^*$ exchange
}\label{decay1}\end{figure}
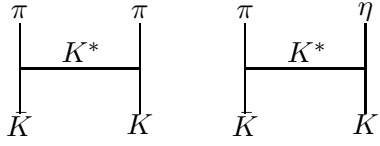

The atomic state of the $K^+K^-$ system can decay through the
strong interaction of exchanging $K^*$ to $\pi\pi$ and $\pi\eta$.
In the following we calculate the decay width from the Feymman
diagram depicted in Fig. \ref{decay1}. The interaction Lagrangians
for the $K^*$ coupling to two pseudoscalar mesons can be written
as,
\begin{equation}\mathcal L_{\pi KK^*}=iG_V\{(\partial_\mu\bar K)\vec \tau
K^{*\mu}\cdot\vec\pi-\bar K\vec \tau
K^{*\mu}\cdot(\partial_\mu\vec\pi)+\bar K^{*\mu}\vec \tau
K\cdot(\partial_\mu\vec\pi)-\bar K^{*\mu}\vec \tau (\partial_\mu
K)\cdot\vec\pi\}\end{equation}
\begin{equation}\mathcal L_{\eta KK^*}=\sqrt3iG_V\left[\partial_\mu\eta(\bar K^{*\mu}
K-\bar KK^{*\mu})+\eta(\partial_\mu\bar KK^{*\mu}-\bar
K^{*\mu}\partial_\mu K)\right] . \end{equation}

With $G_V=g_{\rho\pi\pi}/2$, the coupling constants $g_{\pi KK^*}$
and $g_{\eta KK^*}$ can be determined by the $g_{\rho\pi\pi}$:
\begin{equation}g_{\pi
KK^*}=G_V~~,~~g_{\eta KK^\ast}=-\sqrt3G_V.
\end{equation}

The decay width is:
\begin{equation}
d\Gamma=\frac{1}{32\pi^2}\frac{|{\vec
p}|}{M^2}|\mathcal{M}_I(|B\rangle\to |PP\rangle|^2d\Omega,
\end{equation}

with
$$\mathcal{M}_I(|B\rangle\to |PP\rangle)
=\sqrt{\frac{M}{2m_K^2}}\int d^3r{d^3k\over (2\pi)^3} e^{-i{\vec
k}\cdot{\vec r}}{\psi}({\vec r}) \mathcal{M}_I({\vec k},-{\vec k}\to
|PP\rangle).$$

Here $\psi({\vec r})$ is the wave function of the bound state and
$\vec p$ is the momentum of a meson in the final state in the center
of mass system. $I$ is the isospin.

For the decay to $\pi\pi$, the isospin of $\pi\pi$ is zero. The
decay amplitude is
\begin{eqnarray}
\mathcal M_0(|B>\to|\pi\pi>) &\!=\!& -{1\over
2}\sqrt{\frac{3}{\pi}} G_V^2\sqrt{\frac M{2m_K^2}}\int
drR_0(r)e^{-\Sigma r} \left[4\left(\frac1{r}+\Sigma
\right)\cos(pr)
+\right. \nonumber\\
&&\left.\left( 2(E^2+E^2_f)+4p^2-m_{K^*}^2+{(m_K^2-m_\pi^2)^2\over
m_{K^*}^2}
-\frac4{r^2}-\frac{4\Sigma}{r}\right){\sin(pr)\over p}\right]. \nonumber\\
\end{eqnarray}
Here $\Sigma$ is defined as
$$\Sigma=\sqrt{m_{K^*}^2-(E-E_f)^2},$$
where E is the energy of the K, $E_f$ is the energy of the final
pion. $R_0(r)$ is the radial wave function of the $K^+K^-$
$s$-wave bound state.

We first neglect the influence of the strong interaction to the
Coulomb binding energy and wave function. We find that the decay
width to $\pi\pi$ is
\begin{equation}
\Gamma(K^+K^-\to\pi\pi)= 0.951~\rm{eV}.
\end{equation}
By taking into account the strong interaction,the decay width of
the $K^+K^-$ bound state to $\pi\pi$ becomes
\begin{equation}
\Gamma(K^+K^-\to\pi\pi)=48.4~\rm{eV}.
\end{equation}

In a similar way one can evaluate the amplitude of the
$K^+K^-\to\eta\pi^0$ decay as
\begin{equation}
\begin{array}{rl}\mathcal M(|B\rangle\to
|PP\rangle)=& \displaystyle \frac1{2}\sqrt{\frac{3}{\pi}}
G_V^2\sqrt{\frac M{2m_K^2}}\int
drR_0(r)r\left\{\left[4\left(\frac1{r^2}+\frac{\Sigma_1}
r\right)\cos
pr+\left(-\frac4{pr^3}-\right.\right.\right.\frac{4\Sigma_1}{pr^2}\cr\noalign{\vskip2truemm}
&\displaystyle+\frac{(E+E_\pi)(E+E_\eta)-(E-E_\pi)^2+4p^2-m_{K^*}^2
}{pr}\\
&\displaystyle\left.\left.+\frac{(m_K^2-m_\eta^2)(m_K^2-m_\pi^2)/m_{K^*}^2}{pr}\right)\sin
pr\right]e^{-\Sigma_1
r}\cr\noalign{\vskip2truemm}&\displaystyle+\left[4\left(\frac1{r^2}+\frac{\Sigma_2}
r\right)\cos
pr+\left(-\frac4{pr^3}-\right.\right.\frac{4\Sigma_2}{pr^2}\cr\noalign{\vskip2truemm}
&\displaystyle+\frac{(E+E_\pi)(E+E_\eta)-(E-E_\eta)^2+4p^2-m_{K^*}^2}{pr}\cr\noalign{\vskip2truemm}
&\displaystyle\left.\left.\left.+\frac{
(m_K^2-m_\eta^2)(m_K^2-m_\pi^2)/m_{K^*}^2}{pr}\right)\times\sin
pr\right]e^{-\Sigma_2 r}\right\},\end{array}
\end{equation}
where
$$\Sigma_1=\sqrt{m_{K^*}^2-(E-E_\pi)^2}~~,~~\Sigma_2=\sqrt{m_{K^*}^2-(E-E_\eta)^2},$$
with $E_\pi$ and $E_\eta$ as the energies of the the final $\pi$
and $\eta$ mesons, respectively.

Using the $K^+K^-$ Coulomb wave function the calculated strong
decay width to $\eta\pi^0$ is:
$$\Gamma(K^+K^-\to\eta\pi^0)=0.859~\rm{eV}$$
Taking into account the strong correction, $K^+K^-$ bound state
decay width to $\eta\pi^0$ becomes
$$\Gamma(K^+K^-\to\eta\pi^0)=29.8~\rm{eV}$$
One can see that the decay widths are very small, so the $K^+K^-$
atomic bound state cannot have a large mixing with the $f_0(980)$
or $a_0(980)$.

\begin{figure}
\begin{center}
\includegraphics[width=9cm,height=7cm]{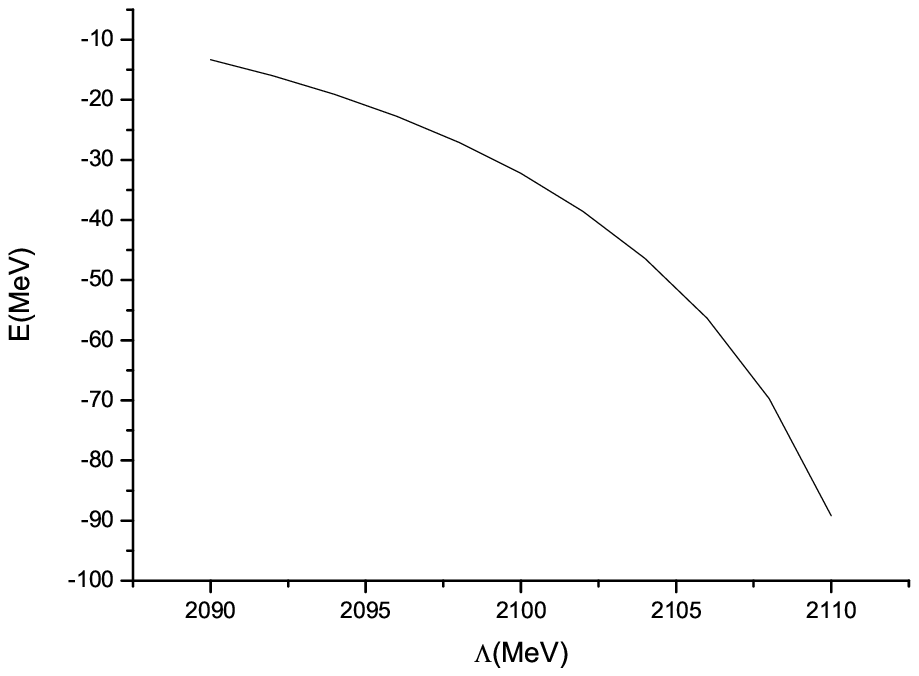}
\caption{Dependence of the $DK$ binding energy on the cutoff
parameter $\Lambda$}\label{fg1}
\end{center}
\end{figure}

Calculations for other two-pseudoscalar-meson systems are
performed with various cutoff parameters $\Lambda$ up to 3.0 GeV.
The dependence of the binding energies of the I=0 systems $DB$,
$D\bar D$, $B\bar B$, $B\bar K$, $B\bar D$, $D\bar K$ and $DK$ on
the choice of the cutoff $\Lambda$ is given in table \ref{tb1} and
Fig. \ref{fg1}. We find that $DB$, $D\bar D$, $B\bar B$ can form
possible s-wave bound states with $\Lambda$ smaller than 2.0 GeV.
The $BK$ cannot form a bound state with $\Lambda<3.0$ GeV. For
two-pseudoscalar-meson systems with isospin I=1, we fail to find
any bound state. For the $KK$, $DD$ and $BB$ pairs, the Bose
symmetry demands their orbital s-wave states to have isospin I=1,
hence no bound states are found.

\begin{table}[h]
\begin{center}\begin{tabular}{c|ccccccc}
\hline \backslashbox{$\Lambda$(GeV)}{E(MeV)}&& DB & $D\bar D$ &
$B\bar B$ & $B\bar K$ & $B\bar D$ & $D\bar K$\\ \hline\hline
1.4&&-1.2&-&-&-&-&-\\
1.5&&-5.7&-&-&-&-&-\\
1.6&&-13.4&-&-&-&-&-\\
1.7&&-24.3&-&-&-&-&-\\ \hline
1.8&&-38.1&-9.3&-0.8&-&-&-\\
1.9&&-54.6&-17.7&-4.3&-&-&-\\
2.0&&-73.6&-28.7&-10.9&-&-&-\\
2.1&&-95.0&-42.4&-21.0&-&-&-\\ \hline
2.2&&-118.6&-58.8&-35.1&-1.4&-&-\\
2.3&&-144.2&-77.9&-53.8&-3.1&-&-\\
2.4&&-171.8&-100.0&-78.2&-5.4&-&-\\\hline
2.5&&-201.3&-125.1&-110.1&-8.2&-0.3&-5.8\\
2.6&&-232.7&-153.7&-152.4&-11.6&-2.5&-55.2\\
2.7&&-265.8&-186.0&-210.3&-15.5&-6.9&-302.7\\
2.8&&-300.6&-222.8&-295.9&-19.9&-13.9&\\
2.9&&-337.2&-264.8&-450.9&-24.7&-24.3&\\
3.0&&-375.6&-313.3&&-30.0&-39.0&\\ \hline\hline
\end{tabular}
\end{center}
\caption{Dependence of binding energies on the cutoff parameter
$\Lambda$ for various isoscalar two-pseudoscalar-meson
systems}\label{tb1}
\end{table}

The $BK$ and $D\bar K$ have the same isospin factor and coupling
constant. One may feel puzzled why the BK cannot form bound state
while $D\bar K$ can. The reason is that generally as the reduce
mass increases the binding energy of two particles becomes larger.
Although the $B$ meson is heavier than $D$ meson, the reduce mass
of the $D\bar K$ is bigger than that of $BK$. So we find that the
$D\bar K$ bound state is possible with $\Lambda\sim 2.5$ GeV,
while the $BK$ system cannot form a bound state.

However the reduce mass of the system is not the only factor to
determine the order of binding energy. In the Hamiltonian given by
Eq.(9), $c(r)/a(r)$ is proportional to something like
$f(m_1,m_2)=(m_1^2+m_2^2+m_1m_2)/(m_1^2+m_2^2+2m_1m_2)$. One finds
$f(m_D,m_B)>f(m_D,m_K)>f(m_B,m_K)$. This factor also influences
the binding energy and may be a reason for the unclear pattern of
the binding energy dependence on the mass of two-meson system
shown in Table \ref{tb1}.

When extending the value of $\Lambda$ to be a little larger than
2.0 GeV, we find that the $DK$ system can also form a bound state.
When $\Lambda>2.07$ GeV, the $DK$ binding energy increases very
rapidly as shown in Fig. \ref{fg1}. If we assume that the
$D_{s0}(2317)$ is a $DK$ molecular state, using the masses of
$m_D=1869$ MeV, $M_K=493$ MeV, the binding energy of $DK$ is 45
MeV. One finds that $\Lambda\approx 2.103$ GeV is needed in order
to generate the $DK$ bound state. In other word, in this model,
the $D_s(2317)$ may be interpreted as a $DK$ bound state .

\section{Summary and discussion}

We use the vector-meson-exchange potential between two
pseudoscalar mesons to search for possible bound states of the
$K\bar K$, $DK$, $B\bar K$, $D\bar D$, $B\bar B$, $BD$, $\bar D
K$, $BK$ and $B\bar D$ systems. We find that the isoscalar $DB$,
$D\bar D$ and $B\bar B$ are most likely to form S-wave bound
states via strong interactions. The isoscalar $DK$ and $B\bar K$
systems are also likely to form S-wave bound state if the cutoff
$\Lambda$ could be larger than 2 GeV. An interesting issue on
these possible two-meson bound states is that all of them can mix
with heavy-quark-antiquark configuration as well as
diquark-antidiquark configuration. For example, the isoscalar $DK$
system has a quark content of $c\bar dd\bar s$ which can transit
to the $c\bar s$ configuration by $d\bar d$ annihilation into a
gluon and can form diquark-antidiquark configuration
$[cd][\overline{ds}]$ by quark re-arrangement. In fact, between
the spin zero diquark $[cd]$ and antidiquark $[\overline{ds}]$,
the vector-meson-exchange force is the same as between $D$ and $K$
mesons. In addition there is an extra color confinement potential
between them. So it is most likely that the $D_{s0}(2317)$ is a
mixture of $c\bar s$, $[cd][\overline{ds}]$ and $DK$
configurations, with $[cd][\overline{ds}]$ component larger than
$DK$ component. For the $c\bar s$ configuration, the $c$ and $\bar
s$ are in relative $P$-wave. For the $[cd][\overline{ds}]$, the
$[cd]$ and $[\overline{ds}]$ are in relative S-wave. It is
possible that instead of having a quark excited to P-wave, the
system prefers to drag out a $q\bar q$ to make a
diquark-antidiquark in relative S-wave. Then the $D_{s0}(2317)$
could be a dominantly 4-quark state as suggested by Cheng and Hou
\cite{Hou} following an earlier idea on the charm-strange 4-quark
state by Lipkin \cite{Lipkin}. A similar scenario happens also in
the baryon sector. For the supposed lowest orbital $l=1$ excited
nucleon state $N^*(1535)$, its properties suggest that it may have
very large $[ud][us]\bar s$ pentaquark components with diquarks
and $\bar s$ all in orbital S-wave \cite{liubc}.

If the cutoff parameter $\Lambda$ could be as large as 2.5 GeV,
then the $B\bar D$ and $D\bar K$ may also form bound states. These
two systems have quark content of $\bar bud\bar c$ and $c\bar
d\bar us$, respectively. So they cannot mix with any ordinary
quark-antiquark configuration. For the corresponding
diquark-antidiquark configurations, $[ud][\overline{bc}]$ and
$[cs][\overline{ud}]$, there is no light-vector-meson-exchange
force, but there is color confinement force. Which configuration
has lower energy needs further investigation.

Other two-pseudoscalar-meson systems including all isovector ones
cannot be bound by the strong interaction sector alone with
$\Lambda<3$ GeV.

For the $K\bar K$ system, the calculation by Ref.\cite{1} suggests
that the $f_0(980)$ may be a $K\bar K$ molecular state bound by
the vector-meson-exchange force. In their calculation, the
momentum dependent terms in the interactions were dropped and a
very large value ($\sim 5.7$ GeV) was assumed for the cutoff
parameter $\Lambda$. We find with momentum dependent terms
included, the $K\bar K$ system cannot be bound by the
vector-meson-exchange interaction. The $f_0(980)$ is more likely a
dominantly $([us][\overline{us}]+[ds][\overline{ds}])/\sqrt{2}$
4-quark state as suggested by Jaffe \cite{Jaffe}. For this kind of
diquark-antidiquark configuration, the vector-meson-exchange force
is the same as for the $K\bar K$ configuration, meanwhile there is
an additional color confinement force to bind them.

With the Coulomb potential, the $K^+K^-$ system can form an atomic
bound state - the kaonium. We find that the binding energy of the
kaonium is 6.58 keV. The binding energy of the kaonium is changed
to 7.05 keV by including the vector-meson-exchange potential. The
decay widths of the kaonium to $\pi\pi$ and $\pi\eta$ are
evaluated.  It is shown that the decay widths of the $K^+K^-$
atomic bound state to $\pi\pi$ and $\eta\pi$ are, respectively,
$$\Gamma(K^+K^-\to\pi\pi)=48.4~\rm{eV}~,
~\Gamma(K^+K^-\to\eta\pi^0)=29.8~\rm{eV}.$$

One can see that the decay widths are very small. We expect that
the mixing of the $K^+K^-$ atomic bound state  with the $f_0(980)$
or $a_0(980)$ is small.

\bigskip
We thank Professor S. Krewald for useful discussion. This work is
partly supported by the National Nature Science Foundation of
China under grants Nos. 10225525, 10435080 and 10475089.


\end{document}